# Time-varying Multivariate Statistical Process Control for Solar Photovoltaic Monitoring and Fault Detecting & Diagnosing Systems

Bundit Boonkhao[1,*] Tararat Mothayakul[2] Chanida Yubolsai[2] and Pornpimol Kavansu[2]

[1]Division of Industrial Engineering and Management, Faculty of Engineering, Nakhon Phanom University, Nakhon Phanom, Thailand, E-mail: bundit@npu.ac.th
[2]Research & Development Institute, Nakhon Phanom University, Nakhon Phanom, Thailand

**Abstract**

Time-varying multivariate statistical process control (TMSPC) has been proposed as a tool for process monitoring, fault detecting & diagnosing of time-varying system. It is a modification of multivariate statistical process control (MSPC) of which is designed for the process that variables operated on normal operational condition which independence of time. However, in some processes, such as solar photovoltaic system, the process variables, i.e., temperature, voltage and current, are time-varying. Therefore, TMSPC has been proposed as monitoring and diagnosing tool for time-varying process. The proposed technique has been demonstrated with solar photovoltaic system located at Research & Development Institute, Nakhon Phanom University, Thailand (RDI-NPU).

## 1. Introduction

Solar energy has been increasingly adopted on a wide scale in various electrical equipment, such as houses, farms, and industries [1]. This is due to the dramatic decrease in the costs of fabrication, installation, and operation. However, despite the cost decrease, the implementation of solar energy is still limited to countries that have a high intensity of sunlight. Moreover, it is also limited during seasons with low light, such as rainy or winter seasons. This implies that the effective application of solar energy highly depends on environment.

Light is an important variable in solar energy production. Normally, photons from light reaching the solar cell panel are not consistent throughout the day. If the day is bright, energy production will be high. However, on foggy or cloudy days, energy production will be lower. This demonstrates that energy production from solar energy is not consistent and depends on the environment surrounding the installation area. Fluctuations in light cause fluctuations in energy production, resulting in poor battery charging performance. This presents challenges for energy management. To manage energy effectively, it is necessary to monitor and analysis the process for further energy planning and operating. Additionally, other environmental variables around the solar cell installation site should also be measured and monitored.

For efficiently monitoring and diagnosing the process with number of variables, multivariate statistical process control (MSPC) should be implemented [2]. The technique implements feature extraction techniques for selecting the most variance variables, then using them for constructing monitoring charts and applying contribution plots for fault diagnosis. The implementation of MSPC is limit for multivariable processes operating under normal condition. This means the process set point is fixed over time. However, solar photovoltaic systems are the time-varying process. The implementation of MSPC for such processes need to be adapted for time-varying processes. Therefore, in this article, time-varying MSPC (TMSPC) has been proposed and demonstrated with solar photovoltaic system.

## 2. Time-varying Multivariate Statistical Process Control

Like traditional MSPC, the procedure of TMSPC is composed of data preparation, training process, monitoring charts and fault detection & diagnosis. This section will briefly describe the proposed TMSPC.

### 2.1 Data preparation

For the collected variables of solar cell data, the dataset is designed as three-dimensional (3D) data **X**[$I \times J \times K$] which $I$ is the date of collection, $J$ is the variable and $K$ is series of time. This can be represented as a block shown in Figure 1. The dataset must be standardised prior to process dimension reduction. Then this dataset will be extracted feature by using principal component analysis (PCA) technique [3] which will be described in the next section.

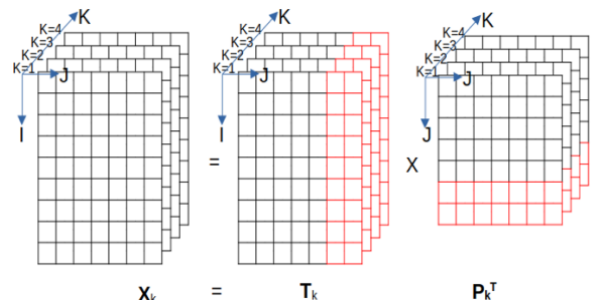

Figure 1 Block diagram of dataset for PCA processing.

## 2.2 Time-varying Principal Component Analysis

PCA can only process the two-dimensional (2D) dataset therefore it is necessary to transform the 3D dataset to 2D dataset. In this article, the reduction of dimension is proposed by slicing dataset at each time. Therefore, the dataset $\mathbf{X}[I \times J \times K]$ will be transformed to $\mathbf{X}_K[I \times J]$, and $\mathbf{X}_K$ refers to data with $[I \times J]$ dimension at time $K$, this is called time-varying PCA (TPCA). The next step is the procedure of PCA.

PCA is the feature extraction method from multivariable dataset. From normal operational dataset $\mathbf{X}[I \times J]$, it can be decomposed as linear combination

$$\mathbf{X} = \breve{\mathbf{T}}\breve{\mathbf{P}}^T \qquad (1)$$

where $\breve{\mathbf{T}}$ is score matrix $[I \times J]$, $\breve{\mathbf{P}}$ is loading matrix $[J \times J]$ and superscript $T$ is transpose of matrix. By reduction of dimension, number of variables $J$ may be selected only $R$ variables, $R \leq J$. So, the dataset $\mathbf{X}[I \times J]$ can be rewritten as

$$\mathbf{X} = \mathbf{TP}^T + \mathbf{E} \qquad (2)$$

where $\mathbf{T}$ is score matrix $[I \times R]$, $\mathbf{P}$ is loading matrix $[J \times R]$ and $\mathbf{E}$ is error matrix $[I \times J]$. This is a training process to obtain score and loading matrix with reduction size of variable to $R$.

The above process is for 2D dataset so for this proposed each time series will be trained and obtained both score and loading matrix at each time $K$.

$$\mathbf{X}_k = \mathbf{T}_k \mathbf{P}_k^T + \mathbf{E}_k \qquad (3)$$

The next process will be using those parameters to construct the monitoring charts.

## 2.3 Monitoring Charts

When obtaining new data at time $k$, the score of the new data can be obtained by multiplying the loading matrix

$$\boldsymbol{t}_k = \boldsymbol{x}_k \mathbf{P}_k \qquad (4)$$

where $\boldsymbol{x}_k$ vector data $[1 \times J]$ at time $k$ and $\boldsymbol{t}_k$ vector score $[1 \times R]$ at time $k$. Monitoring chart can be constructed from the new data, in here, Hotelling's $T^2$ ($T_k^2$) has been constructed.

$$T_k^2 = \sum_{r=1}^{R} \frac{t_{k,r}^2}{s_{t_{k,r}}^2} \qquad (5)$$

where $t_{k,r}^2$ – square score at index $r$ and $s_{t_{k,r}}^2$ – variant of score at index $r$ at time $k$.

The control limit for Hotelling's $T^2$ monitoring chart can be represented as upper control limit (UCL)

$$T_{UCL}^2 = \frac{(I-1)(I+1)R}{I(I-R)} F_v(R, I-R) \qquad (6)$$

where $T_{UCL}^2$ is upper control limit of $T_k^2$, $F_v(\cdot,\cdot)$ is $F$ – distribution.

## 2.4 Fault detection & diagnosis

In the event of $T_k^2$ is greater than $T_{UCL}^2$, this indicates that the system has detected a fault or is out of control. To determine which variables have caused the fault, this can be diagnosed using a contribution plot [4] [5]. Returning to Eq. (4), the contribution plot represents the projection of individual variables onto the loading matrix to generate the scores of individual variables. Therefore, the variable that produces the highest score is the root cause of the process fault. To aid visual comprehension, the contribution plot can be represented using a bar chart illustrating the scores of individual variables.

## 3. Case Study

### 3.1 Solar Photovoltaic System at RDI

To demonstrate TMSPC, a solar photovoltaic system at RDI has been used as a case study. Figure 2 shows the diagram of the solar photovoltaic system installed at RDI. The solar cell system comprises 30 crystalline type solar panels (6 panels connected serially and 5 serial panels connected in parallel, SPOT model S-170-24). The panels are connected to an inverter (LEONICS model G-304). The converted electricity from the inverter has been distributed to the RDI building. Moreover, the building also utilises electricity from the Provincial Electricity Authority (PEA). During operation, seven variables are collected every second: DC voltage [V], DC current [A], AC voltage [V], AC current [A], humidity [%], temperature [°C], and electricity from PEA [V]. The data is stored as a log file on the data logger (DX2000-Wisco Industrial Instruments) and it can also be explored using universal viewer software (SMARTDAC+STANDARD).

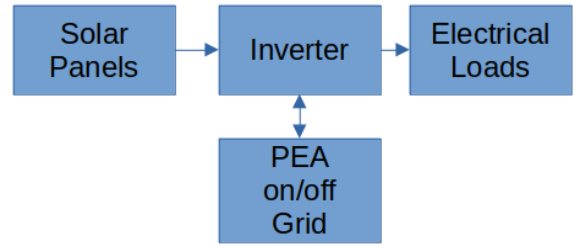

Figure 2 Diagram of solar photovoltaic system install at RDI.

### 3.2 TMSPC for solar system

In this demonstration, the collection of data for 17 days were use as dataset therefore the dimension of data should be $\mathbf{X}[17 \times 7 \times 86400]$, which means 17 days, 7 variables and 86,400 sec. However, the data were divided for training and verifying as 70:30 or 12 days for training dataset and the rest for verifying. Hence, the training dataset were $\mathbf{X}[12 \times 7 \times 86400]$. Figure 3 shows the plots of individual variable of training dataset (12 days) in 86,400 sec. Note that the 12 days were selected from the best normal operational condition (NOC). For TMSPC, the dataset was transformed to $\mathbf{X}_{86400}[12 \times 7]$. In the pre-processing, data $\mathbf{X}_{86400}[12 \times 7]$ was standardised by subtracting with mean and dividing with standard deviation of individual variable. Then, the TPCA was further processed to determine score and loading matrix of the training dataset.

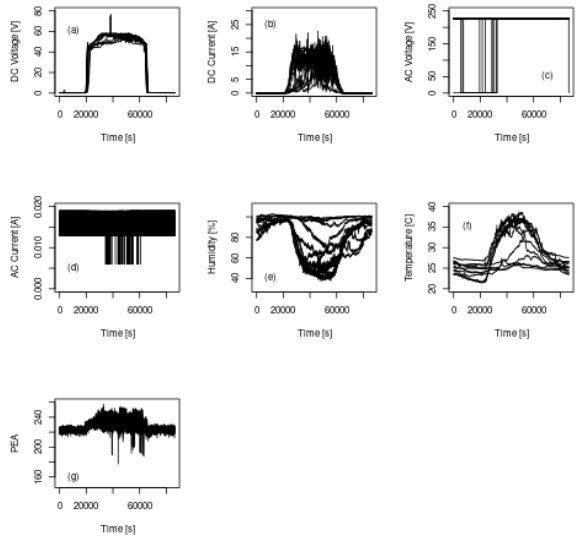

Figure 3 Dataset for training which composed of 12 days and 7 variables: (a) DC voltage, (b) DC current, (c) AC voltage, (d) AC current, (e) humidity, (f) temperature and (g) voltage from PEA.

The score and loading matrix for individual time in second could be calculated using TPCA in Eq. (3). Figure 4(a) represents cumulative scores of principal component (PC) versus time. As can be seen that the first PC (PC1) can be seen to incorporate information from the trained data at least 77.27%. Therefore, only one PC should be enough for creating monitoring charts or selecting $R = 1$. In here, it means that instead of monitoring 7 variables, monitoring only one variable (first PC) can cover at least 77.27% of information.

Figure 4(b) represents $T_k^2$ monitoring charts for the trained data. Seven variables were reduced to only one PC, which was used for constructing the chart. Moreover, by using the upper control limit with a 95% confidence limit for the $T_{UCL}^2$ charts, the variables stay below the $T_{UCL}^2$. This implies that the trained data has sufficient quality for further prediction.

### 3.3 Fault detection & diagnosis

The rest of data remaining was used for evaluating the performance of TMSPC. Firstly, fault detection was evaluated. When the new data obtained, seven variables at individual time $k$ were standardised and the scores of the new data were calculated using Eq. (4) and constructing $T_k^2$ from Eq. (5). Figure 5(a) represents the $T_k^2$ of all 5 new data. As can be seen in the figure that, there were some events that $T_k^2$ higher than $T_{UCL}^2$. This means the faults were detected.

When the fault was detected, fault diagnosing is consequent process for evaluation. By considering individual dataset, as can be seen in Figure 5(b) for only **data15**, there were faults detected along the daytime. To diagnose the root cause of the fault, i.e., which variable caused the fault, this can be determined by using contribution plot. By picking the time that $T_k^2 > T_{UCL}^2$, for example at time 47,619 sec, $T_{47619}^2 = 46.65 > T_{UCL}^2 = 5.24$, then calculating contribution plot following from Eq. (4).

Figure 6 represents contribution plot of the selected times. Figure 6(a) shows the contribution plot while the system was on the NOC, score of each variable is very small and the sum of square of score is less than $T_{UCL}^2$.

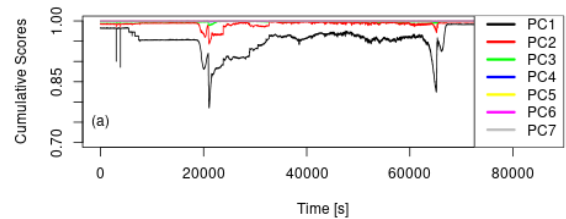
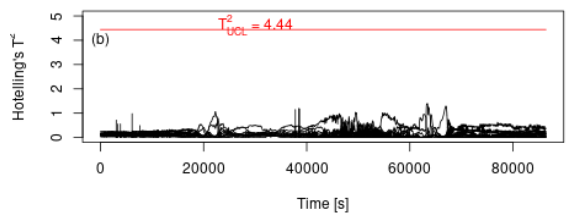

Figure 4 TPCA processed of the training dataset; (a) score of individual principal component along the time, (b) transformation of the training dataset from 7 variables to only one principal component.

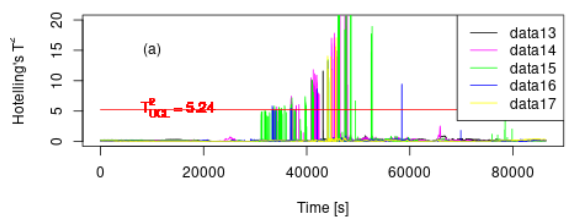
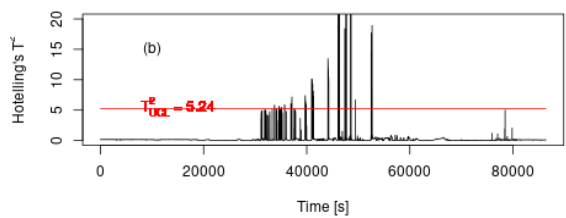

Figure 5 Hotelling's $T^2$; (a) 5 new data, (b) selected at **data15**.

Contribution plot while out of NOC, score of the root cause will be large. At time $k = 47,619$ sec, the fault was detected and, if considering Figure 6(b), the variable DC voltage contributed the highest score. This means DC voltage variable was the root cause of the fault at time 47,619 sec. This was the same for all other faults.

Returning to the original variable **data15** at time 47,619 sec or $\mathbf{X}_{47619}$, Figure 7 shows graphical presentation of all seven variables versus time. As can be seen in Figure 7(a), DC voltage at time 47,619 sec was fluctuated therefore this should be the reason of fault occurred. This is agreed with the diagnosing using contribution plot. Although, DC current variable also fluctuated, the cause was less effect than DC voltage and it was the second most effected. The knowledge for this technique may be used as further maintenance or energy management.

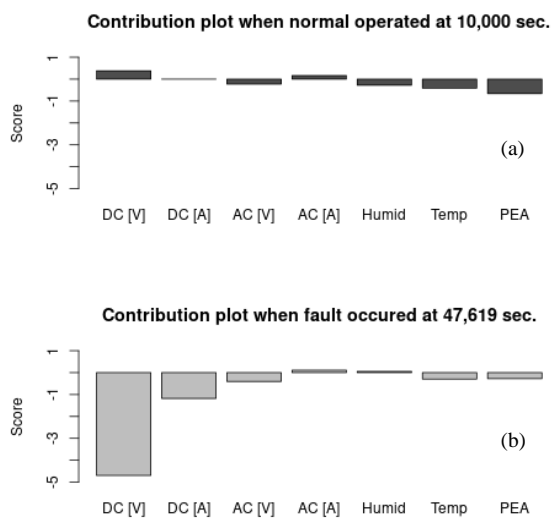

Figure 6 Contribution plot of all variables; (a) when Hotelling's $T_k^2$ less than $T_{UCL}^2$, (b) when Hotelling's $T_k^2$ greater than $T_{UCL}^2$.

## 4. Conclusion

TMSPC has been proposed as a tool for monitoring and fault & diagnosing solar photovoltaic system which is time-varying process. The tool using TPCA for dimension reduction of seven solar photovoltaic variables to only one principal component. Then, Hotelling's $T^2$ and contribution plot were used as monitoring chart and diagnosing, respectively. Both can detect and diagnose faults that occurred from the system at individual time. This technique may apply for further applications such as chemical batch process which the operating condition is time-varying.

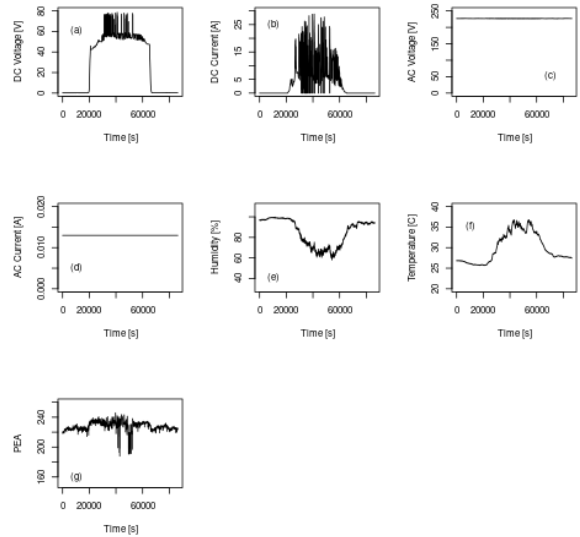

Figure 7 All 7 variables when fault occurred at time 47,619.

## Acknowledgement

This research was financially supported by National Research Council Thailand (NRCT) under the Community Renewable Energy Center (CREC) project grant number 2562A13402012.